\documentclass{article}
\usepackage{newcent}
\usepackage{bifchaos}
\usepackage{multicol}
\usepackage{vmargin}
\usepackage[dvips]{graphicx}

\setpapersize{A4}
\setmarginsrb{1.3cm}{2cm}{2cm}{2.5cm}{0pt}{0mm}{0pt}{0mm}

\begin{document}

\title{\bf DYNAMICS OF ELASTIC EXCITABLE MEDIA}

\author{
JULYAN H. E. CARTWRIGHT\thanks{Email julyan@hp1.uib.es, WWW
http://formentor.uib.es/$\sim$julyan}, 
\\
{\it Instituto Andaluz de Ciencias de la Tierra, IACT (CSIC-UGR),} \\
{\it E-18071 Granada, Spain} \\
\\
V\'ICTOR M. EGU\'ILUZ\thanks{Email victor@imedea.uib.es, WWW
http://www.imedea.uib.es/$\sim$victor}, 
EMILIO HERN\'ANDEZ-GARC\'IA\thanks{Email dfsehg4@ps.uib.es, WWW
http://www.imedea.uib.es/$\sim$emilio}
\& 
ORESTE PIRO\thanks{Email piro@imedea.uib.es, WWW
http://www.imedea.uib.es/$\sim$piro} 
\\
{\it Institut Mediterrani d'Estudis Avan\c{c}ats, IMEDEA (CSIC--UIB),} \\
{\it E-07071 Palma de Mallorca, Spain} \\
}

\date{Int. J. Bifurcation and Chaos, to appear (1999)}

\maketitle

\begin{abstract}
The Burridge--Knopoff model of earthquake faults with viscous friction is
equivalent to a van der Pol--FitzHugh--Nagumo model for excitable media with
elastic coupling. The lubricated creep--slip friction law we use in the
Burridge--Knopoff model describes the frictional sliding dynamics of a range of
real materials. Low-dimensional structures including synchronized oscillations
and propagating fronts are dominant, in agreement with the results of laboratory
friction experiments. Here we explore the dynamics of fronts in elastic
excitable media.
\end{abstract}

\begin{multicols}{2}

\section{Introduction}

\begin{figure*}
\begin{center}
\includegraphics[width=0.6\textwidth]{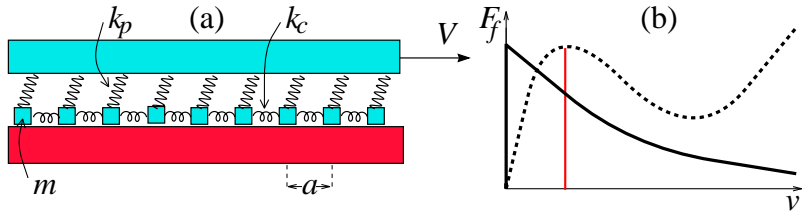}
\end{center}
\caption{\label{friction} (a) The Burridge--Knopoff model. 
(b) The velocity weakening stick--slip friction law of Carlson \& Langer
[$F_f(v)=F_0\,{\mathrm sgn}(v)/(1+|v|)$, where $v$ is the velocity of 
the block] (solid line) and the Burridge--Knopoff type creep--slip friction 
law we use (dashed line), showing the threshold (vertical line) where the block 
starts to slip.
}
\end{figure*}

The Burridge--Knopoff model \cite{burridge} mimics the interaction of two
plates in a geological fault as a chain of blocks elastically coupled together
and to one of the plates, and subject to a friction force by the surface of
the other plate, such that they perform stick--slip motions ---
Fig.~\ref{friction}a. This simple 
system reproduces some statistical features of real earthquakes
\cite{carlson} such as the Gutenberg--Richter power-law distribution
\cite{gutenberg}, considered an example self-organized criticality,
\cite{cls} which obviously involves a large number of degrees
of freedom. However, recent laboratory experiments \cite{rubio} that attempt
to reproduce these dynamics in a real stick--slip dynamical system consisting
of an elastic gel sliding around a metallic cylinder, have shown that
low-dimensional phenomena are more robust in reality, and, although proven to
be unstable in the Burridge--Knopoff model, do show up in the laboratory.

In a different realm, excitable media are usually studied using the model of
van der Pol, FitzHugh, and Nagumo \cite{vdp,fitz1,fitz2,nagumo}. This model
normally includes only diffusive coupling. Originally from physiology and
chemistry, excitable media have also captured the attention of physicists
and mathematicians working in the area of nonlinear science because of the
apparent universality of many features of their complex spatiotemporal
properties \cite{meron}. 

We have shown \cite{quakeletter} that a Burridge--Knopoff model with a
lubricated creep--slip friction force law showing viscous properties at both
the low and high velocity limits (Fig.~\ref{friction}b) is a type of van der
Pol--FitzHugh--Nagumo excitable medium in which the local interaction is
elastic rather than diffusive. We have investigated the dynamics of the model 
and have shown that its behaviour is dominated by low-dimensional structures, 
including global oscillations and propagating fronts. Here we 
investigate further the dynamics of elastic excitable media, and focus on the 
behaviour of the fronts.

\section{The Model}
Our elastic excitable medium model may be written \cite{quakeletter}
\begin{equation}
\frac{\partial^2\chi}{\partial t^2}=c^2\frac{\partial^2\chi}{\partial x^2}
-(\chi-\nu t)-\gamma\,\phi\left(\frac{\partial\chi}{\partial t}\right),
\label{continuum}\end{equation}
where, in the language of frictional sliding, $\chi(x,t)$ represents the
time-dependent local longitudinal deformation of the surface of the upper
plate in the static reference frame of the lower plate,
$\phi(\partial\chi/\partial t)=
(\partial\chi/\partial t)^3/3-\partial\chi/\partial t$ is the friction 
function, as the dashed line in Fig.~\ref{friction}b, $\gamma$ measures the
magnitude of the friction, $c$ is the longitudinal speed of sound, and $\nu$
represents the pulling velocity or slip rate.
Compare this with the discrete Burridge--Knopoff model from whence
Eq.~(\ref{continuum}) may be derived in the continuum limit
\begin{equation}
m \frac{d^2x_i}{dt^2}=
k_c(x_{i+1}-2 x_i+x_{i-1})-k_{p}(x_i-Vt)-F_f\left(\frac{dx_i}{dt}\right),
\label{discrete}\end{equation}
where $x_i$ is the departure of block $i$ from its equilibrium position.  It
has been noted \cite{cls} that in some cases discrete Burridge--Knopoff
models fail to attain a well-defined continuum limit. This is not a
consequence of any numerical instability in computer simulations, but of  the
hyperbolic nature of the equation and of the shape of the friction force
commonly used. In such cases, Eq.\ (\ref{continuum}) should be considered as a
symbolic representation of the well-defined discrete dynamics of Eq.\
(\ref{discrete}). From Eq.\ (\ref{continuum}) we can obtain an expression for
the local velocity $\psi=\partial\chi/\partial t$ of the interface that gives
us the model written as a couple of differential equations of first order in
time
\begin{eqnarray}
\frac{\partial\psi}{\partial t}&=&\gamma(\eta-\phi(\psi)), 
\label{eq1} \\
\frac{\partial\eta}{\partial t}&=&
-\frac{1}{\gamma}\left(\psi-\nu-c^2\frac{\partial^2\psi}{\partial x^2}\right)
\label{eq2}
.\end{eqnarray}\label{eq}

\section{Front Dynamics}

\begin{figure*}
\begin{center}
\includegraphics[angle=-90,width=\textwidth]{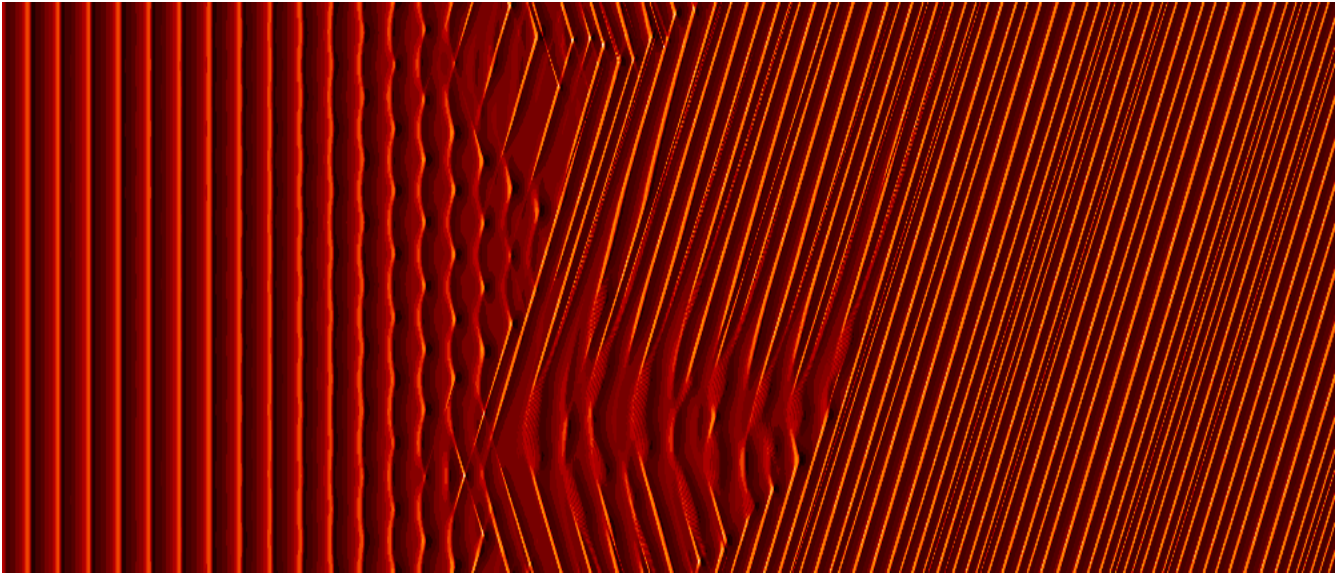}
\end{center}
\caption{\label{pic} 
Evolution of global oscillations into propagating fronts via period doubling; 
a spatiotemporal plot of $\psi(x,t)$, time is horizontal left to right and 
space is vertical.
}
\end{figure*}

The dynamical behaviour of the elastic excitable medium model has been reported
in detail in a previous paper \cite{quakeletter}.  It is notable that much of 
the chaotic behaviour shown by the discrete versions of the Burridge--Knopoff
models built on the basis of a monotonic velocity weakening friction law
\cite{rice,svr,xu}
becomes more organized in our case. Global oscillations and propagating fronts
dominate a large proportion of the relevant parameter space. Moreover, the
global oscillations show interesting instability mechanisms leading to the
appearance of the propagating fronts. Among these instabilities, the most
interesting is perhaps the occurrence of a period-doubling bifurcation at a
finite spatial wavelength. As a result of this bifurcation, the globally
synchronized oscillatory medium breaks into a finite number of equidistant
pacemaker zones from which pairs of counterpropagating fronts are emitted.
Fronts emitted from neighbouring pacemakers annihilate upon collision and the
annihilation point then becomes an emitting centre for the next generation of
fronts; the whole process repeats after two iterations. These results are
graphically summarized by Fig.~\ref{pic}, where the transient evolution of a
nearly uniform initial state is shown for $\nu$ just above the period-doubling
instability. In the first stages of the evolution the dynamics is dominated by
synchronized global oscillations. The instability then grows, giving rise
during a certain time interval to a transient period-doubled structure.
Finally this structure decays into a set of propagating fronts (or propagating
pulses, since a pair of neighbouring fronts can be considered a pulse). This
bifurcation was discussed in more detail in a previous work \cite{quakeletter}.

Here we focus on the properties of the propagating front regime with special
emphasis on the selection mechanisms for the front velocity and spatial
configuration. We suppose a solution of the type
$\psi(x,t)=f(\tilde z)$, where $\tilde z=x/v+t$, and $v$ is the 
front velocity. This together with the further rescaling
$z=\tilde z/\sqrt{1-c^2/v^2}$ leads to 
\begin{equation}
\frac{d^2f}{dz^2}+\mu(f^2-1)\frac{df}{dz}+f=\nu
,\label{fronts}\end{equation}
which is the van der Pol equation with the nonlinearity rescaled by
$\mu=\gamma/\sqrt{1-c^2/v^2}$. The propagating fronts are then periodic
solutions of the van der Pol equation. The parameter $\mu$ is undefined until 
the value of the front velocity $v$ is chosen. However, we know that the 
period of the solution is a function $T=T(\mu)$ of $\mu$: in the limit of 
large $\mu$, $T$ behaves as $T=k\mu+O(\mu^{-1})$, where 
$k=3+(\nu^2-1)\ln[(4-\nu^2)/(1-\nu^2)]$ \cite{fgpv}. Since this period should
be commensurate with the system size $S$, we have the condition
$nT(\mu(v))=S/(v\sqrt{1-c^2/v^2})$, where $n$ is an integer, to select the
allowed front velocities, which in the large $\mu$ limit gives us the
quantizing condition $v=S/(nk\gamma)$. The integer $n$ can be interpreted as
the total  number of pulses propagating in the system. Because Eq.\
(\ref{fronts}) has bounded solutions only if $v^2>c^2$, the propagating fronts
are supersonic. Since our analytical estimations predict that $v$ becomes
smaller than $c$ and approaches zero as the number of pulses in the system is
increased, there should be a maximum number of pulses allowed in the system.  
However, numerical simulations show that it is possible to find solutions
composed of an arbitrary number of propagating pulses with a proper
choice of the initial conditions. A question then arises as to what is the long
term behaviour of these solutions when the number of pulses exceeds the maximum
allowed by the restrictions on the front velocity. There are three simple
scenarios logically compatible with the analysis:
\begin{enumerate}
\item Some of the pulses are annihilated by the dynamics.
\item Since the configuration cannot propagate rigidly at a compatible uniform
speed, nondecaying fluctuations of the velocity of individual pulses
should be observed.
\item The system evolves into a uniformly propagating solution 
with speed approaching the singular limit $v=c$. In this limit the pulses 
become discontinuous, so that Eq.\ (\ref{continuum}) becomes ill-defined.  
The original system of blocks and springs is thus no longer well represented by
the partial differential equation Eq.\ (\ref{continuum}), 
to which our analytic estimations pertain, but the discrete 
effects present in Eq.\ (\ref{discrete}) prevail.
\end{enumerate}

\begin{figure*}
\begin{center}
\includegraphics[angle=90,width=0.6\textwidth]{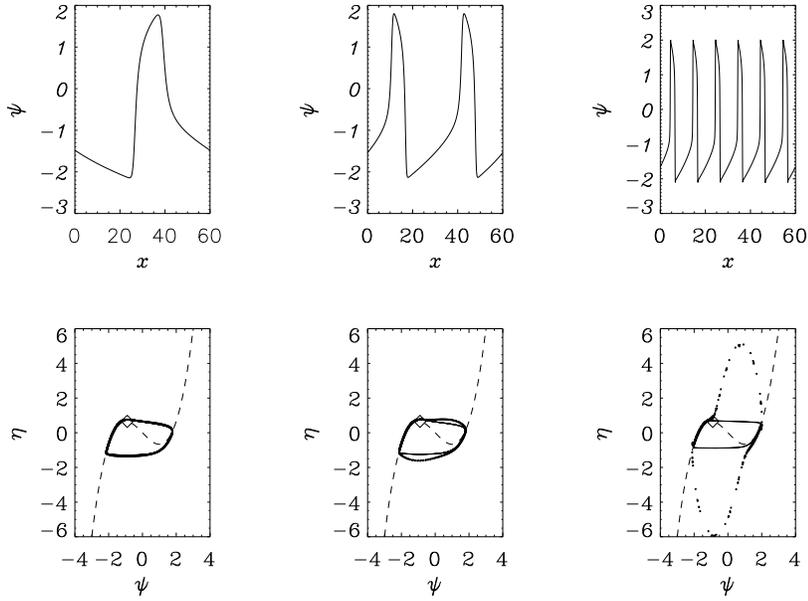}
\end{center}
\caption{\label{pulses}
Changing the initial conditions one can obtain configurations with different
numbers of pulses. In the first row we show $\psi$ against $x$, and in the
second $\eta$ against $\psi$. In the second row the points are data from 
numerical integration, the continuous line is the limit cycle of Eq.\
(\ref{fronts}) with $\mu$ calculated from the parameters and the observed
$v$, the dashed line is the nullcline, and the diamond shows the position of
the unstable fixed point.
First column: 
1 pulse (2 fronts), velocity $v=4.78\pm 0.64$; has the form of the 
van der Pol--FitzHugh--Nagumo limit cycle.
Second colum: 2 pulses (4 fronts), velocity $v=2.47\pm 0.35$; shows small 
deviations from the van der Pol--FitzHugh--Nagumo continuum limit. 
Third column: 6 pulses (12 fronts), velocity $v=1.10\pm 0.06$; is  
clearly outside the continuum limit.
The solitary pulse in the first row is travelling to the right; the others to
the left. Note from the velocity of the solitary pulse that there cannot be
five or more pulses in the continuum limit: since $v\propto 1/n$, they would 
then have a subsonic velocity. Parameters are $\gamma=3$, $\nu=-0.9$,
$c=1$, system size $L=60$, discretization $dx=60/1024$. Integration is by
fourth-order Runge--Kutta in time, and the second-order finite differences
implied by Eq.\ (\ref{discrete}) in space. 
}
\end{figure*}

\begin{figure*}
\begin{center}
\includegraphics[width=\textwidth]{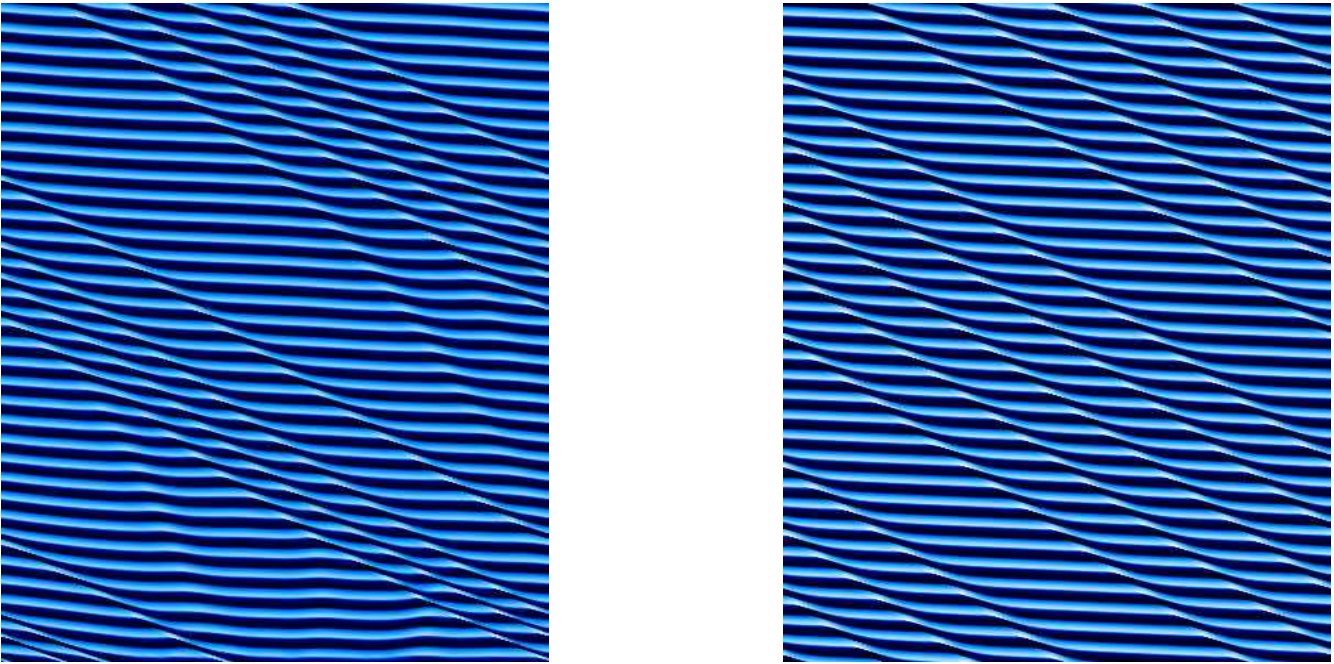}
\end{center}
\caption{\label{vic}
The formation of phase fronts between regions of synchronous 
oscillations; spatiotemporal plots of $\psi(x,t)$, 
time is vertical and space is horizontal. 
}
\end{figure*}

The first scenario occurs, but only during a transient phase: at long times
states with arbitrary numbers of pulses are reached and maintained in time.
Examples of how the second scenario could be realized will be briefly discussed
below. We first focus on the third scenario, for which we find abundant
numerical evidence. Figure~\ref{pulses} summarizes this. In upper frame of the
first column we see a propagating solution composed of only one pulse
travelling around the system to the right. Using the spatial coordinate $x$ as
a parameter we have drawn below the phase-space diagram of such a solution.
This diagram shows geometrically that the spatial dependence of the solution is
accurately described as the periodic solution of a van der
Pol--FitzHugh--Nagumo equation. The second column shows a two-pulse solution
propagating to the left at a speed approximately one half that of the 
single-pulse solution, and thus closer to the sound velocity $c$, as
expected from the analytical estimations. The phase-space projection below
shows general agreement with the prediction of Eq.\ (\ref{fronts}), although a
small discrepancy is already visible. Finally, the third column in
Fig.~\ref{pulses} shows a breakdown of the continuum model. A train of six
pulses travelling at constant speed has been generated, which is forbidden
within the framework of the continuum analysis. They travel to the left at
speed essentially $c$. Notice the sharp gradients at both the leading and
trailing edges of the pulses. If we refine the numerical spatial 
discretization, this just increases the gradients, so that we may expect
discontinuities in the continuum limit. The discrete model Eq.\ (\ref{discrete})
introduced in the computer should better represent the physical system of 
springs and blocks than the partial differential equation Eq.\ 
(\ref{continuum}). The phase-space diagram clearly shows a strong departure
from the solutions of the van der Pol--FitzHugh--Nagumo equations. 
A well-defined phase-space structure appears, which should 
be understood as a property of the purely discrete system, different 
in this case from the van der Pol--FitzHugh--Nagumo phase-space structure. 
Discreteness effects have been studied in several models with velocity 
weakening friction \cite{rice,svr,xu,galeano}.

When $\nu$ has been taken very far from the slipping threshold the behaviour
locally is of autonomous relaxation oscillations, and a slightly different kind
of propagating front dynamics occurs. By setting $\nu = 0$ we recover the 
symmetric form of the van der Pol equation, supplemented in our
case with the elastic spatial term. Now a Floquet stability analysis of the
homogeneous oscillatory state shows the development of a long wavelength
instability as $\nu$ decreases. The evolution of this instability is displayed
in Fig.~\ref{vic}. One can roughly describe this behaviour as the formation of
spatial regions where the medium undergoes almost synchronous relaxation
oscillations. Neighbouring regions, however, oscillate in antiphase and a
phase-change front defining the border between them travels around the system.
On the left we can see some of these fronts moving at different velocities
during a transient stage, while on the right is displayed an asymptotic state
where the fronts have reached an equilibrium configuration in which the phase
distribution travels rigidly. This behaviour is reminiscent of the phase
dynamics found in the complex Ginzburg--Landau equation 
\cite{montagne,montagne2}.

A closer look at Fig.~\ref{vic} indicates that the spatial regions that we
describe as synchronized, are in fact regions where the excitation or 
relaxation propagates extremely fast. This is visualized in the picture as a
very slight tilt of the bands representing the oscillation. Notice that the
projection of these bands shrinks considerably near the places where the phase
jumps. By taking spacelike slices of these pictures we obtain the instantaneous
configuration of the excitation field. Such configurations look like travelling
pulses varying wildly in size and velocity. This behaviour can be considered an
extreme manifestation of scenario 2 above.

\section{Discussion}
A lubricated friction law in the Burridge--Knopoff model can be justified both
by theory \cite{persson,persson2}, and experiments
\cite{heslot,brechet,kilgore,demirel,budakian}.
Moreover, studies of peeling adhesive tape \cite{hong}, of Saffman--Taylor
fracture in viscous fingering \cite{kurtze}, and of the Portevin--Le
Ch\^atelier effect \cite{kubin,lebyodkin} lead to the same form of friction law
as we use here. Certainly our model well represents the qualitative
characteristics of the laboratory stick--slip dynamics experiments 
\cite{rubio} referred to above, and might have relevance to the present debate 
on shear stress and friction in real geological faults
\cite{sleep,melosh,cohen}. In the context of excitable systems elastic coupling
has been left aside, because in the chemical and biological systems studied up
to now the coupling is diffusive. However, an elastic excitable medium can be
realized as an active transmission line or optical waveguide \cite{quakeletter}.

\section*{Acknowledgements}
The authors acknowledge the financial support of the Spanish Direcci\'on 
General de Investigaci\'on Cient\'\i fica y T\'ecnica, contracts PB94-1167 and 
PB94-1172.

\bibliographystyle{bifchaos} 
\bibliography{database}

\begin{thebibliography}{}

\bibitem[\protect\citeauthoryear{Brechet \& Estrin}{1994}]{brechet}
Brechet, Y. \& Estrin, Y. [1994]
\newblock ``The effect of strain rate sensitivity on dynamic friction of
  metals,''
\newblock {\em Scripta Metall.} {\bf 30}, 1449--1454.

\bibitem[\protect\citeauthoryear{Budakian \bgroup \em et al.\egroup
  }{1998}]{budakian}
Budakian, R., Weninger, K., Hiller, R.~A. \& Putterman, S.~J. [1998]
\newblock ``Picosecond discharges and stick--slip friction at a moving meniscus
  of mercury on glass,''
\newblock {\em Nature} {\bf 391}, 266--268.

\bibitem[\protect\citeauthoryear{Burridge \& Knopoff}{1967}]{burridge}
Burridge, R. \& Knopoff, L. [1967]
\newblock ``Model and theoretical seismicity,''
\newblock {\em Bull. Seismol. Soc. Am.} {\bf 57}, 341--371.

\bibitem[\protect\citeauthoryear{Carlson \& Langer}{1989}]{carlson}
Carlson, J.~M. \& Langer, J.~S. [1989]
\newblock ``Properties of earthquakes generated by fault dynamics,''
\newblock {\em Phys. Rev. Lett.} {\bf 62}, 2632--2635.

\bibitem[\protect\citeauthoryear{Carlson \bgroup \em et al.\egroup
  }{1994}]{cls}
Carlson, J.~M., Langer, J.~S. \& Shaw, B.~E. [1994]
\newblock ``Dynamics of earthquake faults,''
\newblock {\em Rev. Mod. Phys.} {\bf 66}, 657--670.

\bibitem[\protect\citeauthoryear{Cartwright \bgroup \em et al.\egroup
  }{1997}]{quakeletter}
Cartwright, J. H.~E., Hern\'andez-Garc\'{\i}a, E. \& Piro, O. [1997]
\newblock ``{Burridge-Knopoff} models as elastic excitable media,''
\newblock {\em Phys. Rev. Lett.} {\bf 79}, 527--530.

\bibitem[\protect\citeauthoryear{Cohen}{1996}]{cohen}
Cohen, P. [1996]
\newblock ``Inside the {San Andreas},''
\newblock {\em New Scientist} {\bf 2017}, 24--27.

\bibitem[\protect\citeauthoryear{Demirel \& Granick}{1996}]{demirel}
Demirel, A.~L. \& Granick, S. [1996]
\newblock ``Friction fluctuations and friction memory in stick--slip motion,''
\newblock {\em Phys. Rev. Lett.} {\bf 77}, 4330--4333.

\bibitem[\protect\citeauthoryear{Feingold \bgroup \em et al.\egroup
  }{1988}]{fgpv}
Feingold, M., Gonz\'alez, D.~L., Piro, O. \& Viturro, H. [1988]
\newblock ``Phase locking, period doubling, and chaotic phenomena in externally
  driven excitable systems,''
\newblock {\em Phys. Rev. A} {\bf 37}, 4060--4063.

\bibitem[\protect\citeauthoryear{FitzHugh}{1960}]{fitz1}
FitzHugh, R.~A. [1960]
\newblock ``Thresholds and plateaus in the {Hodgkin--Huxley} nerve equations,''
\newblock {\em J. Gen. Physiol.} {\bf 43}, 867--896.

\bibitem[\protect\citeauthoryear{FitzHugh}{1961}]{fitz2}
FitzHugh, R.~A. [1961]
\newblock ``Impulses and physiological states in theoretical models of nerve
  membrane,''
\newblock {\em Biophys. J.} {\bf 1}, 445--466.

\bibitem[\protect\citeauthoryear{Galeano \bgroup \em et al.\egroup
  }{1998}]{galeano}
Galeano, J., Espa{\~n}ol, P. \& Rubio, M.~A. [1998]
\newblock ``Dynamics of solitary relaxations in stick--slip,''
\newblock {\em preprint}.

\bibitem[\protect\citeauthoryear{Gutenberg \& Richter}{1956}]{gutenberg}
Gutenberg, B. \& Richter, C.~F. [1956]
\newblock ``Magnitude and energy of earthquakes,''
\newblock {\em Ann. Geofis.} {\bf 9}, 1--15.

\bibitem[\protect\citeauthoryear{Heslot \bgroup \em et al.\egroup
  }{1994}]{heslot}
Heslot, F., Baumberger, T., Perrin, B., Caroli, B. \& Caroli, C. [1994]
\newblock ``Creep, stick--slip, and dry friction dynamics: Experiments and a
  heuristic model,''
\newblock {\em Phys. Rev. E} {\bf 49}, 4973--4988.

\bibitem[\protect\citeauthoryear{Hong \& Yue}{1995}]{hong}
Hong, D.~C. \& Yue, S. [1995]
\newblock ``Deterministic chaos in failure dynamics: Dynamics of peeling in
  adhesive tape,''
\newblock {\em Phys. Rev. Lett.} {\bf 74}, 254--257.

\bibitem[\protect\citeauthoryear{Kilgore \bgroup \em et al.\egroup
  }{1993}]{kilgore}
Kilgore, B.~D., Blanpied, M.~L. \& Dieterich, J.~H. [1993]
\newblock ``Velocity dependent friction of granite over a wide range of
  conditions,''
\newblock {\em Geophys. Res. Lett.} {\bf 20}, 903--906.

\bibitem[\protect\citeauthoryear{Kubin \& Estrin}{1985}]{kubin}
Kubin, L.~P. \& Estrin, Y. [1985]
\newblock ``The {Portevin}--{Le Ch\^atelier} effect in deformation with
  constant stress rate,''
\newblock {\em Acta Metall.} {\bf 33}, 397--407.

\bibitem[\protect\citeauthoryear{Kurtze \& Hong}{1993}]{kurtze}
Kurtze, D.~A. \& Hong, D.~C. [1993]
\newblock ``Tip dynamics in {Saffman--Taylor} fracture,''
\newblock {\em Phys. Rev. Lett.} {\bf 71}, 847--850.

\bibitem[\protect\citeauthoryear{Lebyodkin \bgroup \em et al.\egroup
  }{1995}]{lebyodkin}
Lebyodkin, M.~A., Brechet, Y., Estrin, Y. \& Kubin, L.~P. [1995]
\newblock ``Statistics of the catastrophic slip events in the {Portevin}--{Le
  Ch\^atelier} effect,''
\newblock {\em Phys. Rev. Lett.} {\bf 74}, 4758--4761.

\bibitem[\protect\citeauthoryear{Melosh}{1996}]{melosh}
Melosh, H.~J. [1996]
\newblock ``Dynamical weakening of faults by acoustic fluidization,''
\newblock {\em Nature} {\bf 379}, 601--606.

\bibitem[\protect\citeauthoryear{Meron}{1992}]{meron}
Meron, E. [1992]
\newblock ``Pattern formation in excitable media,''
\newblock {\em Phys. Rep.} {\bf 218}, 1--66.

\bibitem[\protect\citeauthoryear{Montagne \bgroup \em et al.\egroup
  }{1997}]{montagne2}
Montagne, R., Hern{\'a}ndez-Garc{\'{\i}}a, E., Amengual, A. \& {San Miguel}, M.
  [1997]
\newblock ``Wound-up phase turbulence in the complex {Ginzburg--Landau}
  equation,''
\newblock {\em Phys. Rev. E} {\bf 56}, 151--167.

\bibitem[\protect\citeauthoryear{Montagne \bgroup \em et al.\egroup
  }{1996}]{montagne}
Montagne, R., Hern{\'a}ndez-Garc{\'{\i}}a, E. \& {San Miguel}, M. [1996]
\newblock ``Winding number instability in the phase-turbulence regime of the
  complex {Ginzburg--Landau} equation,''
\newblock {\em Phys. Rev. Lett.} {\bf 77}, 267--270.

\bibitem[\protect\citeauthoryear{Nagumo \bgroup \em et al.\egroup
  }{1962}]{nagumo}
Nagumo, J.~S., Arimoto, S. \& Yoshizawa, S. [1962]
\newblock ``An active pulse transmission line simulating nerve axon,''
\newblock {\em Proc. IREE Aust.} {\bf 50}, 2061--2070.

\bibitem[\protect\citeauthoryear{Persson}{1995}]{persson}
Persson, B. N.~J. [1995]
\newblock ``Theory of friction: Stress domains, relaxation, and creep,''
\newblock {\em Phys. Rev. B} {\bf 51}, 13568--13585.

\bibitem[\protect\citeauthoryear{Persson}{1997}]{persson2}
Persson, B. N.~J. [1997]
\newblock {\em Sliding Friction: Physical Principles and Applications}
\newblock (Springer).

\bibitem[\protect\citeauthoryear{Rice}{1993}]{rice}
Rice, J.~R. [1993]
\newblock ``Spatio-temporal complexity of slip on a fault,''
\newblock {\em J. Geophys. Res.} {\bf 98}, 9885--9907.

\bibitem[\protect\citeauthoryear{Rubio \& Galeano}{1994}]{rubio}
Rubio, M.~A. \& Galeano, J. [1994]
\newblock ``Stick--slip dynamics in the relaxation of stresses in an elastic
  medium,''
\newblock {\em Phys. Rev. E} {\bf 50}, 1000--1004.

\bibitem[\protect\citeauthoryear{Schmittbuhl \bgroup \em et al.\egroup
  }{1993}]{svr}
Schmittbuhl, J., Vilotte, J.-P. \& Roux, S. [1993]
\newblock ``Propagatve macrodislocation modes in an earthquake fault model,''
\newblock {\em Europhys. Lett.} {\bf 21}, 375--380.

\bibitem[\protect\citeauthoryear{Sleep \& Blanpied}{1992}]{sleep}
Sleep, N.~H. \& Blanpied, M.~L. [1992]
\newblock ``Creep, dislocation and the weak rheology of major faults,''
\newblock {\em Nature} {\bf 359}, 687--692.

\bibitem[\protect\citeauthoryear{van~der Pol \& van~der Mark}{1928}]{vdp}
van~der Pol, B. \& van~der Mark, J. [1928]
\newblock ``The heart beat considered as a relaxation oscillator and an
  electrical model of the heart,''
\newblock {\em Phil. Mag. (7)} {\bf 6}, 763--775.

\bibitem[\protect\citeauthoryear{Xu \& Knopoff}{1994}]{xu}
Xu, H.-J. \& Knopoff, L. [1994]
\newblock ``Periodicity and chaos in a one-dimensional dynamical model of
  earthquakes,''
\newblock {\em Phys. Rev. E} {\bf 50}, 3577--3581.

\end{thebibliography}

\end{multicols}

\end{document}